\documentclass[aps,reprint,superscriptaddress]{revtex4-2}

\usepackage{graphicx}
\usepackage{amssymb}   
\usepackage{amsmath}    
\usepackage{graphicx}   
\usepackage{verbatim}   
\usepackage{color}      
\usepackage{hyperref}   
\usepackage{bm} 
\usepackage{ulem}

\begin{document}

\title{Third-order nonlinear transport in a percolative two-dimensional superconductor} 

\author{ Wenjun Liu}
\altaffiliation{These authors contributed equally to this work.}
\affiliation{Wuhan National High Magnetic Field Center and School of Physics, Hubei Fundamental Research Center for Physics, Hubei Key Laboratory of Gravitation and Quantum Physics, Huazhong University of Science and Technology, Wuhan 430074, China}

\author{ Chenghe Wang}
\altaffiliation{These authors contributed equally to this work.}
\affiliation{Wuhan National High Magnetic Field Center and School of Physics, Hubei Fundamental Research Center for Physics, Hubei Key Laboratory of Gravitation and Quantum Physics, Huazhong University of Science and Technology, Wuhan 430074, China}

\author{ Xiubin Li}
\affiliation{Wuhan National High Magnetic Field Center and School of Physics, Hubei Fundamental Research Center for Physics, Hubei Key Laboratory of Gravitation and Quantum Physics, Huazhong University of Science and Technology, Wuhan 430074, China}

\author{ Kenji Watanabe}
\affiliation{ Research Center for Electronic and Optical Materials, National Institute for Materials Science, 1-1 Namiki, Tsukuba 305-0044, Japan}

\author{ Takashi Taniguchi}
\affiliation{ Research Center for Materials Nanoarchitectonics, National Institute for Materials Science, 1-1 Namiki, Tsukuba 305-0044, Japan}

\author{ Tao Zhang}
\affiliation{Wuhan National High Magnetic Field Center and School of Physics, Hubei Fundamental Research Center for Physics, Hubei Key Laboratory of Gravitation and Quantum Physics, Huazhong University of Science and Technology, Wuhan 430074, China}

\author{ Xiao-Xiao Zhang}
\email[Contact author: ]{xxzhang@hust.edu.cn}
\affiliation{Wuhan National High Magnetic Field Center and School of Physics, Hubei Fundamental Research Center for Physics, Hubei Key Laboratory of Gravitation and Quantum Physics, Huazhong University of Science and Technology, Wuhan 430074, China}

\author{Jing Li}
\email[Contact author: ]{jing\_li@hust.edu.cn}
\affiliation{Wuhan National High Magnetic Field Center and School of Physics, Hubei Fundamental Research Center for Physics, Hubei Key Laboratory of Gravitation and Quantum Physics, Huazhong University of Science and Technology, Wuhan 430074, China}

\date{\today}

\begin{abstract}
Percolative superconductivity frequently arises in two-dimensional van der Waals materials due to reduced dimensionality, enhanced quantum fluctuations, and complex electron-phonon interactions, providing a unique platform where normal electrons coexist with Cooper pairs. We report the observation of substantial third-order nonlinear transport in a trilayer 1$T^{\prime}$-MoTe$_2$ superconductor within its percolative transition regime. The third-harmonic longitudinal voltage ($V_{\|}^{3\omega}$) exhibits a clear cubic dependence on excitation current below a threshold, with both its magnitude and nonlinear coefficient strongly correlated with the superconducting state. This nonlinear response is semi-quantitatively captured by the superconducting fluctuation within the time-dependent Ginzburg–Landau theory, where nonlinear transport arises due to fluctuating Cooper pairs. Our results demonstrate that third-order nonlinear transport serves as a sensitive probe of superconducting transitions in percolative systems and establish a foundation for exploring higher-order transport phenomena in strongly correlated systems.
\end{abstract}
 
\maketitle
Nonlinear transport provides a powerful probe of underlying electronic properties that go beyond the scope of linear response. In systems without magnetic ions or impurities, Hall signals typically require an external magnetic field. However, nonlinear Hall effects—especially higher-order ones—can emerge even in zero field, revealing mechanisms inaccessible through first-order transport. Among these, second-order nonlinear responses have attracted significant attention. They were theoretically predicted to arise from Berry curvature dipoles in non-centrosymmetric systems with time-reversal symmetry \cite{gao2014field, sodemann2015quantum, du2018band, zhang2018berry,ye2025engineering}, and subsequently observed in topological materials \cite{xu2018electrically, ma2019observation, kang2019nonlinear, wang2024non} and moiré superlattices \cite{he2022graphene, sinha2022berry,cao2025nonlinear}. In addition, giant second-order effects have also been reported in twisted two-dimensional (2D) systems, attributed to impurity and phonon scattering, skew scattering, and other extrinsic mechanisms \cite{duan2022giant, huang2023giant, zhong2024effective}. Third-order nonlinear Hall effects, on the other hand, have been primarily linked to Berry connection polarizability in systems that preserve both inversion and time-reversal symmetry \cite{lai2021third, liu2022berry, wang2022room, ye2022orbital, zhao2023gate}.

Recent advances continue to uncover new mechanisms behind higher-order transport, including second-order responses driven by the quantum metric \cite{wang2023quantum, gao2023quantum,kaplan2024unification}, and higher-order anomalous Hall effects associated with Berry curvature multipoles \cite{sankar2023experimental, zhang2023higher} and quantum metric multipoles \cite{liu2025giant,zhao2025magnetic}. Beyond transverse responses, higher-order longitudinal nonlinearities have also begun to draw attention. While early studies focused on nonreciprocal charge transport in noncentrosymmetric superconductors \cite{wakatsuki2017nonreciprocal, zhang2020nonreciprocal, wu2022nonreciprocal}, more recent work uncovered sizable third-order longitudinal signals in topological chiral antiferromagnetic semimetals, possibly linked to the emergence of charge density wave order \cite{mi2023third,chen2024charge}. Intriguingly, recently observed third-order nonlinear Hall responses in quantum Hall systems have been attributed to strong electron-electron (\textit{e–e}) interactions \cite{he2024third}, highlighting the central role of many-body correlations in nonlinear transport phenomena.

2D van der Waals superconductors, such as atomically thin MoS$_2$ \cite{lu2015evidence, costanzo2016gate, fu2017gated}, NbSe$_2$ \cite{xi2016ising, tsen2016nature}, and TaS$_2$ \cite{navarro2016enhanced, de2018tuning, peng2018disorder}, provide a distinct platform where strong \textit{e–e} interactions can coexist with finite resistance—ideal conditions for exploring correlation-induced nonlinear transport responses. Unlike conventional bulk superconductors, which exhibit a sharp transition at a well-defined critical temperature ($T_c$), these materials often show broad superconducting (SC) transitions, characteristic of percolative behavior. This arises from reduced dimensionality, enhanced quantum fluctuations, disorder, and complex electron-phonon interactions. In this extended transitional regime—where SC correlations develop without a full zero-resistance state—strong \textit{e–e} interactions may play an increasingly important role, offering a promising window into higher-order nonlinear phenomena, reminiscent of recent findings in the quantum Hall regime \cite{he2024third}.

\begin{figure}[!htbp]
    \centering
    \includegraphics[width=0.95\linewidth]{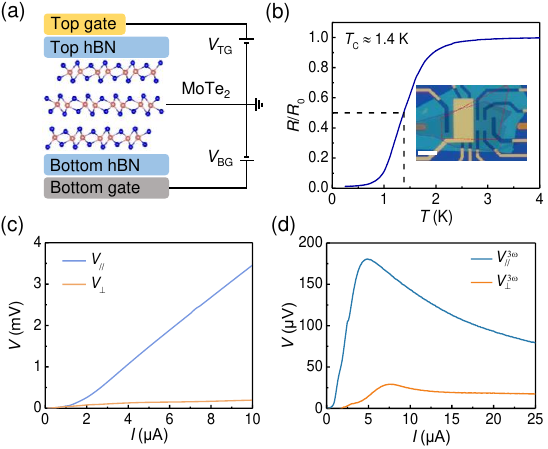}
    \caption{Setup of experiments. (a) Device schematic.
(b) Superconducting transition in $\rm{MoTe_{2}}$ showing the resistance normalized to its value at 4 K. Inset: Optical image of device A, with the dashed red line outlining the $\rm{MoTe_{2}}$ sheet. The scale bar represents 10 $\mu$m.
(c-d) First- and third-harmonic longitudinal voltage ($V_{\|}$) and transverse voltage ($V_{\perp}$) as a function of the longitudinal excitation current ($I$).}
    \label{fig:enter-label}
\end{figure}

Here, we report the observation of third-order nonlinear transport behavior in a percolative SC trilayer 1$T^{\prime}$-$\rm{MoTe_2}$, revealing a close correlation between the nonlinear response and the SC state. This study also demonstrates that the third-order signals can be tuned by carrier density, and suggests that the nonlinear response may originate from tunable fluctuating Cooper pairs at low temperatures. These findings provide an ideal platform for studying higher-order nonlinear transport phenomena in 2D SC systems.

Figure 1(a) shows a dual-gated $\rm{MoTe_2}$ device fabricated using a layer-by-layer dry-transfer technique (see Section I of \cite{SM}). The crystal structure was identified as the 1$T^{\prime}$ phase, with the $T_d$ phase excluded based on temperature-dependent Raman spectroscopy and second-harmonic generation measurements (see Fig. S2 of \cite{SM}). Figure 1(b) presents the temperature-dependent resistance of device A, from which all main-text data were obtained (see inset for optical micrograph). A separate device B was also studied and exhibited qualitatively similar behavior (see Section XIV of \cite{SM}). A clear SC transition is observed at $T < 3$ K, and we determine the subcritical transition temperature $T_c \approx $ 1.4 K as the temperature at 50{\%} of the normal state resistance. Notably, at base temperatures, the resistance does not drop to zero but instead stabilizes around 2 $\Omega$. We refer to these systems as percolative superconductor due to their broad transition range and inability to achieve strictly zero resistance at low temperatures. Full $R$–$T$ data over an extended temperature range are plotted in Fig. S3 of \cite{SM}.

Devices were measured using a standard low-frequency lock-in technique at 177.77 Hz. Higher-order transport data were obtained from the corresponding harmonics of the lock-in signal, with frequency-dependent checks performed for consistency (see Fig. S4(a) of \cite{SM}). All measurements were conducted in a pumped He-3 cryostat at a base temperature of 250 mK, unless stated otherwise. To minimize errors from geometric components of the electrodes, we measured both longitudinal and transverse voltages simultaneously as a function of excitation current $I$ [Fig. 1(c)]. The $I$–$V$ curves remain nearly linear for $I > 4~\mu\text{A}$, where the sample deviates from its SC state (also see Fig. S11 for d$V$/d$I$ measurements). At zero magnetic field, the first-order Hall signal is entirely from the longitudinal voltage, and probe misalignment effect is estimated to be 6{\%} from the ratio of longitudinal to transverse voltages. Figure 1(d) shows the nonlinear dependence of $V_{\|}^{3\omega}$ and $V_{\perp}^{3\omega}$ on $I$, with $V_{\perp}^{3\omega}$ corrected for this geometric component. The longitudinal nonlinear response is substantially larger than in $\text{WTe}_2$ \cite{ye2022orbital} and $\text{CoNb}_3\text{S}_6$ \cite{mi2023third}, while the nonlinear Hall response is comparable to that in quantum Hall systems attributed to strong \textit{e–e} interactions \cite{he2024third}, and greatly exceeds values in other well-studied materials \cite{ye2022orbital,lai2021third,zhao2023gate}. These results indicate a pronounced enhancement of nonlinear transport driven by strong \textit{e–e} interactions.

\begin{figure}[!htbp]
    \centering
    \includegraphics[width=1\linewidth]{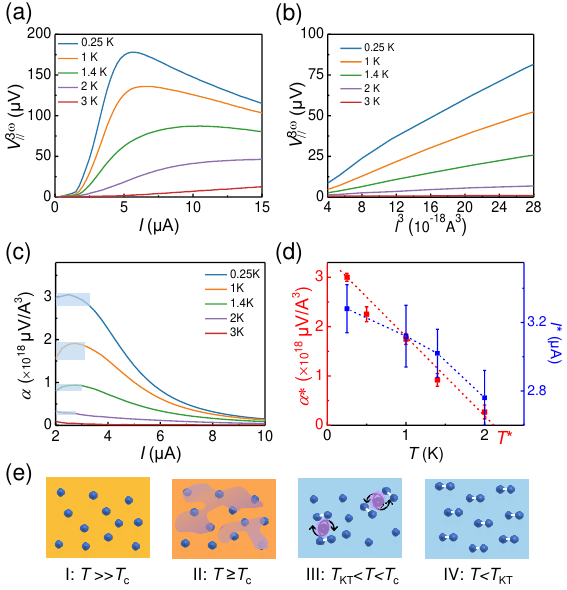}
    \caption{Temperature dependence of third-order nonlinear transport properties. (a) {$V_{\|}^{3\omega}$} versus {$I$} at different temperatures. (b) {$V_{\|}^{3\omega}$} as a function of {$I^{3}$} at different temperatures, showing a linear dependence. (c) {$V_{\|}^{3\omega}$}/{$I^{3}$} as a function of {$I$} at different temperature. The blue shaded boxes mark the regimes where $\alpha$ remains relatively large. The width of each box extends from the lowest presented current up to $I=I^*$, and the height spans the full range of fluctuating $\alpha$ values within that regime. (d) Third-order nonlinear transport coefficients $\rm{\alpha^*}$ and subcritical current scale $I^*$ as functions of temperature. (e) Cartoon illustrating the evolution of superconducting fluctuations varies with temperature. Depending on the temperature, these can be divided into four regions: The normal state (I), the fluctuation state (II), the vortex state (III) and the superconducting state (IV).}
    \label{fig:enter-label}
\end{figure}

To further examine the relationship between the third-order signal and the SC state, we measured the third order $V$-$I$ curves at different temperatures, as shown in Fig. 2(a). The third-order signal becomes more prominent at lower temperatures, reaching its maximum at intermediate current values before diminishing at higher currents. In the low-current regime ($I < \sim  4~\mu$A), the signal strength systematically decreases with increasing temperature and vanishes entirely by $T = 3$ K. This temperature dependence closely matches the range over which the resistance drop occurs, as shown in Fig. 1(b). At higher currents, the third-order signal weakens as the SC state is increasingly suppressed. A similar non-monotonic current dependence of the third-order longitudinal response has been observed in a topological chiral antiferromagnetic semimetal below its Néel temperature, where nonlinearities also appear in the first-order $V$–$I$ characteristics \cite{mi2023third}.

Below the percolative SC transition temperature, marked by a substantial drop in resistance in Fig. 1(b), the third-order signal exhibits a clear cubic dependence on the excitation current in the low-current regime, where superconductivity remains relatively robust, as shown in Fig. 2(b) (the cubic relationship over the full current range is presented in Fig. S5 of \cite{SM}). Notably, the cubic scaling does not extend to zero current because both the first- and third-order longitudinal voltage signals become extremely weak at base temperature under very small excitation currents ($I<\sim1\mu$A), where the sample is considered to be deep in the SC state. This analysis follows the procedure adopted for higher-order transport in the quantum Hall regime \cite{he2024third}. 

To quantitatively characterize the third-order signal, we analyze the ratio $V_{\parallel}^{3\omega} / I^3$, denoted as $\alpha$.  As shown in Fig. 2(c), $\rm{\alpha}$ remains comparably large in the low-current regime, fluctuating around a characteristic value, $\rm{\alpha^*}$, before decreasing rapidly as the excitation current increases further. The characteristic coefficient $\alpha^*$ is obtained from a simple cubic fit to the $V_{\parallel}^{3\omega}$--$I$ data (see Fig.S6 of \cite{SM}). We further define a subcritical current scale, $I^*$, as the maximum current up to which the cubic fit remains valid. Thus, $I^*$ is a fitting parameter that characterizes the current range over which the third-order response follows cubic scaling, rather than an independently defined SC critical current.

The extracted values of $\alpha^*$ and $I^*$ are summarized in Fig.~2(d). The third-order nonlinear coefficient $\alpha^*$ decreases approximately linearly with increasing temperature and extrapolates to zero at a characteristic temperature $T^* \approx 2.1$ K, which closely matches the onset temperature of the percolative SC transition determined independently from the linear transport measurements [Fig.1(b)]. This agreement demonstrates a strong correlation between the third-order nonlinear response and the emergence of superconductivity. Although $I^*$ is not intended to represent the SC critical current, it decreases systematically with increasing temperature, qualitatively following the expected evolution of a SC current scale. The fitting procedure, the definition of $I^*$, and the determination of the associated error bars are described in Fig.S6 of \cite{SM}. Interestingly, the extracted $I^*$ is also close to the current scale at which a peak appears in the differential resistance ($\mathrm{d}V/\mathrm{d}I$) measured at base temperature (Fig.~S11), although we emphasize that $I^*$ is introduced solely as a descriptor of the validity range of the cubic scaling and is not used as an independent measure of superconductivity.

To elucidate the origin of the third-order transport signals, we consider SC fluctuations within the framework of time-dependent Ginzburg–Landau (TDGL) theory (see Section XVI of \cite{SM}). We derive the effective GL free energy from a low-energy $k\cdot p$ model of the present system, which accounts for the order parameter or Cooper pair fluctuation. Generalizing the Aslamazov-Larkin-Schmid paraconductivity\cite{LG1968,schmid1969diamagnetic}, we find when $T_c\lesssim T$ both the linear $\sigma^{(1)}$ and the leading nonlinear effect, i.e., the third order, in the present centrosymmetric system. The latter takes the form $\sigma^{(3)} \propto \epsilon^{-4}$ with $\epsilon = \ln(T/T_c)$ under Gaussian Cooper pair fluctuations. It
leads to a third-order resistivity $\rho^{(3)} = -\tfrac{\sigma^{(3)}}{[\sigma_n + \sigma^{(1)}]^4}$ with $\sigma_n$ the normal conductivity, which is strongly enhanced as $T \to T_c$ and corroborates with the experimental data semi-quantitatively (as shown in Fig. S23). This is achieved, besides experimentally extracted quantities, without artificial parameter fine-tuning. This highly remarkable numerical agreement thus strongly supports the percolative-fluctuation scenario. A qualitative analysis towards broader temperature ranges is briefly discussed in Section XVI of \cite{SM}, while an additional quantitative microscopic construction lies beyond the scope of this Letter. Within this framework, the nonlinear response originates from the propagation of the dynamically fluctuating SC order parameter or Cooper pairs, mediated by their coupling to normal electrons. The broad transition regime in percolative two-dimensional superconductors enables strong SC fluctuations over an extended range, making the third-order response experimentally accessible.

The SC fluctuation can be naturally divided into three regimes. 
(i) For $T \gtrsim T_c$, both amplitude and phase fluctuations of the order parameter are present, the Cooper pairs can transport charges but with a finite lifetime, which dominantly contributes to the third-order transport. 
(ii) For $T_{\rm KT} < T < T_c$ with $T_{\rm KT}$ the Kosterlitz--Thouless transition temperature, the amplitude is well developed while phase fluctuations persist and can form bound and unbound phase topological defects as vortices\cite{RevModPhys.59.1001}. Under a uniform current flow, freely roaming vortices are driven by a Magnus force to induce an electric voltage as per the Josephson relation\cite{hoshino2018nonreciprocal}. However, the current also assists in unbinding the vortex pairs such that the free vortex density depends on the current and hence the generation of nonlinear $\rho^{(3)}$ response. 
(iii) For $T < T_{\rm KT}$, phase coherence is established and there are no thermally unbound vortices. 
The linear response $\rho^{(1)}$, directly contributed by free vortices, hence vanishes; while the current-assisted nonlinear response may still be nonvanishing.
In our experiment, transitions between these regimes are controlled not only by temperature but also by excitation current, which dynamically tunes the strength of SC fluctuations. Figure 2(e) presents a schematic illustration of this process.

This fluctuation-based picture explains the observed non-monotonic current dependence. At base temperature and weak current excitations ($I<\sim1\mu$A), the SC state is well developed and fluctuations are very weak, resulting in negligible third-order response. As the current increases, Cooper pair density and phase rigidity are progressively weakened, overall enhancing the fluctuations and the mechanisms induced by them. This leads to a pronounced third-order signal and the observed maximized values in $\alpha$. When the current is further increased, SC correlation itself is suppressed, reducing the corresponding fluctuations and weakening the nonlinear response. This leads to a rapid decrease in both the magnitude of the third-order signal and the value of $\alpha$, consistent with the breakdown of the cubic scaling between $V_{\parallel}^{3\omega}$ and $I$.

\begin{figure}[!htbp]
    \centering
    \includegraphics[width=1\linewidth]{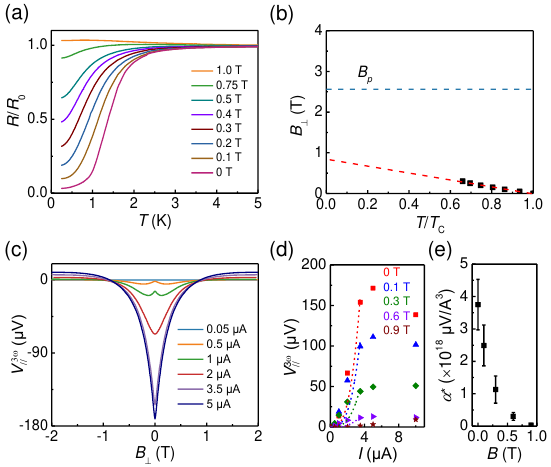}
    \caption{Magnetic field effects on superconductivity and third-order nonlinear transport properties. (a) Temperature-dependent resistance under varied applied perpendicular magnetic fields. (b) Subcritical transition temperature (black symbols), normalized to its zero-field value, as a function of perpendicular magnetic field. The red dashed line represents a fit to the 2D Ginzburg-Landau model, while the blue dashed line indicates the Pauli limit. (c) Magnetic field dependence of the third-order longitudinal voltage {$V_{\parallel}^{3\omega}$} for varied {$I$} at 250 mK. (d) $V_{\parallel}^{3\omega}$ versus $I$ for different magnetic fields. (e) Estimated third-order transport coefficient $\alpha^*$ as a function of magnetic field in the low-current regime ($I < 4~\mu$A). The large error bars reflect deviations from ideal cubic dependence and represent the upper and lower bounds of cubic fits that capture most of the data points. The dashed lines in (d) indicate cubic trends using the $\alpha^*$ values plotted in (e).}
    \label{fig:enter-label}
\end{figure}

Upon applying a perpendicular magnetic field ($B_\perp$), the SC transition is progressively suppressed, evidenced by a reduction in $T_c$ and the eventual disappearance of superconductivity. Figure 3(a) shows temperature-dependent resistance under various $B_\perp$ values, measured with a low excitation current of 50 nA to minimize Joule heating (see also $B_\perp$-dependent resistance at fixed temperatures in Fig. S7(a) of \cite{SM}). The normalized $T_c$ values as a function of $B_\perp$ are plotted in Fig. 3(b), revealing a linear suppression of $T_c$ that extrapolates to zero at $B_{c,\perp} \approx 0.9$ T. This behavior is well captured by the 2D Ginzburg–Landau (GL) relation, $B_{c,\perp} = \frac{\phi_0}{2\pi\xi_{\text{GL}}^2}(1 - T/T_c)$, yielding a GL coherence length $\xi_{\text{GL}} \approx 19.4$ nm. The corresponding BCS coherence length, $\xi_0 = 1.35\xi_{\text{GL}} \approx 26.2$ nm, is significantly larger than the estimated mean free path $l \approx 7.4$ nm, suggesting a percolative SC state with strong fluctuations. This interpretation is supported by the relatively small $B_{c,\perp}$, well below the Pauli limit ($B_p = 1.86T_c \approx 2.6$ T), consistent with previous reports \cite{rhodes2021enhanced, tang2023ambipolar}. In addition, quantum Griffiths singularities observed near the critical field (see Fig. S7 of \cite{SM} for details) also point to significant fluctuation or localization effects influencing the phase transition.

Remarkably, as shown in Fig. 3(c), the amplitude of the third-order longitudinal signal decreases rapidly with increasing $B_\perp$ and vanishes at $B_\perp \approx 0.9$ T, precisely matching the critical field $B_{c,\perp}$. This strong correlation points to a direct connection between the third-order nonlinear response and the SC state. A slight anomalous increase in the third-order signal near $B = 0$ T (for currents $I < 2$ $\mu$A) may arise from increased dwell time due to weak carrier localization, which can enhance interactions among charge carriers and fluctuating Cooper pairs while the Cooper pair density remains largely unaffected. In the low-current regime ($I < 4$ $\mu$A), we empirically fit $V_{\parallel}^{3\omega}$ to a cubic function of $I$ to extract an effective third-order transport coefficient $\alpha^*$ [Fig. 3(e)]. While this fit is not exact---as reflected by the large error bars in Fig. 3(e)---it captures the dominant trend across different $B_\perp$ values, as shown by the dashed lines in Fig. 3(d). The extracted nonlinear coefficient $\alpha^*$ decreases monotonically, but nonlinearly, with increasing $B_\perp$ and vanishes near $0.9$ T [Fig. 3(e)], coinciding with the breakdown of superconductivity. 
This simultaneous disappearance of nonlinear transport and the SC state reinforces the interpretation that the third-order response is governed by SC fluctuations and transient Cooper pairs. Notably, unlike temperature, which tunes $\alpha^*$ approximately linearly, the magnetic field modulates SC fluctuations and vortex dynamics in a distinct, nonlinear manner, leading to a corresponding evolution of the third-order response.

\begin{figure}[!htbp]
    \centering
    \includegraphics[width=1\linewidth]{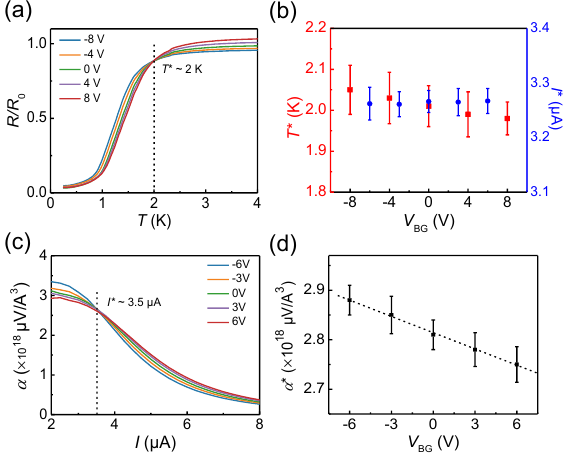}
    \caption{Gate-tunable superconductivity critical transition temperature and nonlinear coefficients. (a) Temperature-dependent $R_{xx}$ at varied back gate voltages, with resistance normalized to its value at 4 K. (b) Characterized percolative SC transition temperature and subcritical current scale at different gate voltages. (c) Ratio of {$V_{\|}^{3\omega}$} to {$I$} as a function of the applied longitudinal excitation current at different gate voltages. (d) Third-order nonlinear transport coefficients $\rm{\alpha^*}$ as a function of back gate voltage.} 
    \label{fig:enter-label}
\end{figure}

We next examine the gate tunability of the third-order transport response. Figure 4(a) shows the temperature-dependent resistance at various back gate voltages ($V_{BG}$), focusing on the SC transition regime. As $V_{BG}$ increases, the onset temperature of the percolative SC transition, $T^*$, which marks the steep resistance drop, shifts to lower values (for the determination of $T^*$, see Fig. S8 of \cite{SM}). This trend is summarized in Fig. 4(b), which plots $T^*$ as a function of $V_{BG}$, indicating a suppression of superconductivity with increasing electron doping. $\rm{MoTe_2}$ is a compensated semimetal hosting both electron and hole pockets near the Fermi level, and previous studies have shown that superconductivity in thin layers of $\rm{MoTe_2}$ is optimized near the carrier compensation point ($\Delta n = 0$) \cite{jindal2023coupled}. Hall measurements reveal an apparently large carrier density of $\sim 1.9 \times 10^{14}$ cm$^{-2}$ at $V_{BG}=0$, with strong gate dependence (see Fig. S12 of \cite{SM}). Given the small gate-induced density change ($\Delta n \sim 5 \times 10^{12}$ cm$^{-2}$), this indicates that the system remains in a compensated semimetallic regime. In this limit, the Hall response is governed by a two-band model \cite{rhodes2021enhanced}, and the extracted Hall carrier density does not reflect the true net electron density, which is expected to be on the order of $\sim 10^{13}$ cm$^{-2}$. The observed gate dependence of $T^*$ therefore reflects the modulation of SC coherence within this compensated regime.

Across all applied $V_{BG}$ values, $V_{\|}^{3\omega}$ retains a cubic dependence on the excitation current in the low-current regime, up to the subcritical current scale $I^*$, which remains nearly constant at 3.2-3.3 $\mu$A (see Fig. S8 of \cite{SM}). Within this regime, the nonlinear coefficient forms local maxima around $\alpha^*$ for each gate voltage, as shown in Fig. 4(c). Although the value of $\alpha^*$ vary weakly with $V_{BG}$, the trend is systematic and non-negligible, as plotted in Fig. 4(d).  This result highlights the sensitivity of third-order transport as a probe of SC coherence and fluctuation, which can be effectively tuned in the percolative transition regime via gate-induced carrier modulation. We note that back-gate doping may also induce a small perpendicular electric field, potentially breaking inversion symmetry in our trilayer $\rm{MoTe_2}$ device. Disentangling the roles of carrier doping and electric field effects will require further study (see Section XIII of \cite{SM}).

Third-order Hall responses were also characterized as functions of excitation current, temperature, and magnetic field (see Fig. S10 of \cite{SM}). These signals also exhibit a clear correspondence with the onset of percolative superconductivity, paralleling the behavior observed in longitudinal measurements. Moreover, thermal effects are known to generate higher-order transport signals in systems with strong temperature-dependent resistance, detailed analysis (see Section IX of \cite{SM}) confirms that Joule heating is not the dominant mechanism behind the observed third-order responses in our system.

In summary, we report pronounced third-order nonlinear transport in the percolative superconducting regime of trilayer 1$T^{\prime}$-MoTe$_2$, with behavior tightly correlated to the superconducting state, and tunable via temperature, excitation current, magnetic field, and gate voltage. The nonlinear response is quantitatively described within a superconducting fluctuation framework, where nonlinear transport arises from the dynamically fluctuating Cooper pair correlation. Importantly, based on experimental input, this theory provides a fine-tuning-free quantitative agreement with experiment across the fluctuation-dominated window ($T_c \lesssim T$)---a highly nontrivial result that strongly supports the percolative-fluctuation scenario. This mechanism does not rely on material-specific band-structure features. Instead, it reflects a general property of two-dimensional percolative superconductors, where an extended fluctuation regime enables a wide and robust window for the interplay between electrons and superconducting correlations. Our results establish third-order nonlinear transport as a sensitive probe of superconducting coherence and fluctuation and their evolution, providing a new route to investigate electron correlation effects in low-dimensional superconductors.

\textit{Acknowledgments}: X.-X.Z. thanks the helpful discussion with W. Yang. Work at the WHMFC was supported by the National Key Research and Development Program of China (grant numbers 2022YFA1602700, 2022YFA1402400), the Natural Science Foundation of China (grant numbers 12274155, 12574172, 52202172), the Knowledge Innovation Program of Wuhan-Basic Research (grant number 2023010201010040) and the Fundamental Research Funds for the Central Universities. K.W. and T.T. acknowledge support from the JSPS KAKENHI (grant numbers 21H05233 and 23H02052) and the World Premier International Research Center Initiative (WPI), MEXT, Japan. The computation was partially completed in the HPC Platform of Huazhong University of Science and Technology.

\end{document}